|
| Keywords | Cold ambient air plasma, water sorption isotherm, dielectric properties, seeds LCR measurements, plasma-seeds interactions |

# 1. Introduction

## 1.1 Context

Faced with growing concerns about biodiversity loss, seeds are now recognized as vital resources. Indeed, they offer potential solutions for ecosystem restoration, sustainable agriculture and forestry, in addition to their pivotal role in global food security [1], [2]. Maintaining seed quality, which encompasses factors such as vigor, controlled dormancy and resistance to pathogens and pests, is therefore of paramount importance. Some conventional farming practices, though effective in various contexts, have been criticized for their reliance on phytosanitary treatments. These treatments, which may include active substances, formulants and pelleting adjuvants, are used to improve germination and seed protection [3]. However, it is essential to differentiate between these substances, as their environmental and health impacts can vary significantly. Some pesticides, particularly older or non-selective ones, have been associated with persistence in the environment and potential health problems [4]. These concerns stem from observations of bioaccumulation in ecosystems and potential adverse effects on non-target species [5], [6]. As our understanding of these challenges grows, increasing emphasis is being placed on finding alternative, sustainable farming methods, including innovative approaches such as cold plasma treatments.

## 1.2. Influence of plasma on seeds

Cold plasmas are weakly ionized gases that can be produced in ambient air by partially ionizing molecular nitrogen, oxygen and water vapor. This leads to the formation of reactive oxygen/nitrogen species (RONS) with lifetimes ranging from a few ms (e.g. OH, NO) to several seconds (e.g. $O_3$).

Cold plasma sources find growing interest in responding to biological challenges, thanks to the possibility of generating a large spectrum of RONS doses and to adapt them to each targeted application. Thus, in the medical field, cold plasma at lower dose can induce cell stimulation/proliferation for wound healing while at higher dose it can drive to senescence and cell death for oncology applications [7]. In agriculture and more specifically in the seed sector, cold plasmas are processed to improve germination parameters (vigor, homogeneity, dormancy release, tolerance to water stress) as well as optimal development of seedlings at early stages [8], [9], [10]. Cold plasmas are also designed to protect seeds from potential fungi (e.g. Fusarium spp., Pythium spp., Tilletia spp.), bacteria (e.g. Xanthomonas, Ralstonia solanacearum) and pests (e.g. P. interpunctella, ants, birds) [11], [12], [13], [14]. Combating these threats is essential as they can compromise agricultural production, nutritional contents and health safety [15]. In addition to all these plasma-triggered biological effects (gene regulation, seed biochemical and physiological changes), surface state modifications are also reported, especially chemical functionalization, texturization and etching of seeds' coatings [16].

## 1.3. Influence of seeds on plasma

To date, most research into plasma seed treatment has focused on the impact of plasma on seeds. Conversely, very little information is available on how seeds can modify the plasma's characteristics. Though this inquiry might initially seem inconsequential, analogous questions have surfaced in other cold plasma domains, especially in the processing of greenhouse gases. Hence, in a packed-bed dielectric barrier device, the efficiency of $CO_2$ splitting significantly depends on the arrangement of the pellets (or beads), and more specifically on their individual size (or bead) [17], [18]. In the case of granulated materials (ZnO catalyst, sea salt NaCl) stacked in a DBD, a transition from the filamentary mode to a combination of filamentary and surface discharges can also be observed [19].

In the specific case of plasma seed treatment, this question is paramount because a same plasma device can be used to treat different volumes of seeds and different types of seeds, i.e. seeds that may change in terms of size, shape, water content, lipid or protein storage. These variables affect both the capacitive and resistive components of the seeds, which in turn can modify the plasma discharge itself. For example, when a dielectric barrier device (DBD) is powered at 8 kV (600 Hz) to process seeds, Judée et al. have shown that a batch of lentil, bean or corn seeds has a capacitance of around 20 pF, which translates into a plasma power of 1350 mW. In contrast, a batch of radish, coriander or sunflower seeds has a capacitance of 14 pF, resulting in a plasma power of 1200 mW [20]. The shape of the seeds can also affect the discharge properties since it determines both the contact areas between seeds as well as the tridimensional spaces (interstices) between them where micro-discharges occur. This physical feature can influence streamers distribution, plasma current and voltage, and therefore the plasma chemical properties. As a result, varying amounts of RONS are produced, which can subsequently impact plasma efficiency [20], [21].

In addition to the influence of seeds' shape and interstices on the discharge properties, the impact of seed water content remains questionable. Even if this issue has not been addressed in the plasma community yet, it appears crucial as it may influence electric field strength and RONS production within the plasma phase.

## 1.4. Understanding the relevance of controlling seed water content in seed treatment processes

The importance of WC in seeds is underscored by its profound impact on seed longevity, i.e. the period in which a seed remains viable, maintaining its ability to germinate and produce a healthy plant. A seed with high longevity can be stored for several years and retain its ability to germinate [22]. Longevity depends on genetic characteristics, exposure to pathogens but also on storage conditions. Among these factors, temperature and humidity control the rubbery or glassy state of cell cytoplasm, the later one being preferred as it reduces seed metabolic activity [23], [24].

Now that the relevance of seed water content has been highlighted, the question of its measurement arises. Usually, seed water content is measured following the oven drying method: a known weight of seeds is dried in an oven at a specific temperature (usually around 105°C) for a predetermined time (typically 24 hours). Then, the seeds are reweighed, and the moisture content is calculated based on the weight loss. If this standard method is very accurate, it is time-consuming, destructive and is therefore not effective to the factory-installed seed treatment systems. Near-infrared reflectance (NIR) spectroscopy and microwave humidimetry are interesting alternatives as well as LCRmetry which is non-intrusive, easy-to-handle and low-cost [25], [26]. Several works illustrate the relevance of LCRmetry:

- In the case of buckwheat seeds packed in a coaxial cylindrical capacitor, Zhu et al. have demonstrated that the dielectric constant ($ε_r'$) and loss factor ($ε_r''$) decrease with rising frequency and increase with rising moisture content (11.1-17.1 %$_{DW}$) and temperature (5°C-40°C) [27]. Their work suggests that LCRmetry can be an effective method to gauge the moisture content of buckwheat.
- Mai and his team have designed a device for measuring seed moisture using a specific type of capacitor [28]. This tool is designed for integration into machines that provide continuous corn drying, facilitating extended storage with minimized spoilage or mildew risks. The device's readings accurately match to actual seed moisture levels, particularly for WC between 14 %$_{DW}$ and 21 %$_{DW}$.
- Moura and his team closely monitored moisture levels during the conversion of sorghum grain to flour, to improve the milling process and the quality of the final product. They identified a clear relationship between sorghum moisture content (13-23 %$_{DW}$) and the ε' and ε'' parameters [29].

Overall, the parameters measured by LCRmetry can be significantly influenced by seed water content [30]. As a result, LCR parameters such as $ε_r'$ and $ε_r''$ can be utilized as data inputs for models designed to estimate seed water content without direct testing. In a second time, they can also be utilized to make predictions about seed viability [27], [28], [29].

## 1.5. Outline

In this article, dormant seeds of Arabidopsis have been treated by a cold ambient air plasma (C2AP) generated in a dielectric barrier device (DBD). We have investigated how seeds water content can modify the electrical properties of plasma, considering four scenarios: no seeds (Control) and three sets of plasma-treated seeds, each with different moisture levels (3 %$_{DW}$, 10 %$_{DW}$ and 30 %$_{DW}$). Then, we demonstrate how C2AP can release dormancy of these Arabidopsis seeds while considering the influence of WC. To decipher the mechanisms ruling the plasma-triggered dormancy alleviation, LCR measurements have been carried out to investigate the bulk dielectric properties of the seeds as well as sorption isotherm measurements correlated with sorption fitting models to identify which of their thermodynamic properties could impact the plasma electrical properties.

# 2. Materials & Methods

## 2.1. Plant material, seed moisture regulation, germination assays

*Arabidopsis thaliana* (ecotype Columbia Col-0) seeds have been harvested dormant following the method described by [31], and then stored at –20 °C to keep dormancy. All seeds have an initial moisture content of 7 %$_{DW}$.

According to the method described by Vertucci and Roos [32], several equilibrium water contents (WC) are obtained by placing Arabidopsis seeds in different hermetically sealed jars, at 20 °C for 4 days. Seeds are placed over silica gel and saturated salt solutions of $ZnCl_2$, $LiCl$, $CaCl_2$, $MgCl_2$, $Ca(NO_3)_2$, $NaCl$, $KCl$ and $KH_2PO_4$, hence resulting to the respective values of relative humidities ($H_{rel}$): 2.5, 9.5, 17.0, 28.5, 40.0, 56.5, 75.0, 83.5 and 93.0 %, respectively. For each condition of relative humidity, three replicates of fresh seeds (approximately 15 mg of seeds per replicate) are dried during 24 h at 105 °C. The mass ratio between water-equilibrated mass to anhydrous mass results in the seed water content. In this work, values of WC are expressed on a dry weight basis (DW) in $g_{H2O} \cdot g_{DW-1}$ as the mean of the three replicates with its standard deviation.

The germination assays are achieved considering four replicates of 100 seeds each, placed on moistened cotton wool in 9 cm diameter glass Petri dishes. These dishes are placed in dark, a condition known to enhance Arabidopsis seed primary dormancy [31]. Seed germination is observed twice daily for 12 days, noting when the seed's radicle protruded the envelope. Each germination curve is obtained by monitoring samples of 400 seeds, with 100 seeds per Petri dish so that the mean values were calculated on 4 × 100 seeds with standard deviation.

## 2.2. Sorption isotherms & Sorption fitting models

*2.2.1. Sorption isotherms*

Seed water properties can be studied by plotting sorption isotherms. These curves describe the relationship between the equilibrium water content of a seed and the relative humidity of the surrounding air at a constant temperature. In simpler terms, they indicate the amount of water a seed absorbs or releases as a function of the humidity of its environment, at a given temperature. A prerequisite for this method is that the seeds must be completely equilibrated with water vapor, which usually takes several days [33].

The experimental datapoints plotted in a WC=f($H_{rel}$) diagram, are generally analyzed using a sorption fitting model (see following sections) that relies on biophysical parameters which are specific to the seed-water interaction but also to the physiological processes at the functional level of seeds. These biological parameters depend also on developmental maturity, desiccation tolerance degree and dormancy state [34].

In this study, four sorption fitting models are evaluated to extract parameters from the experimental data: the Brunauer-Emmett-Teller (BET) model, the Guggenheim-Anderson-de Boer (GAB) model, the D'Arcy and Watt (D&W) model, and the Generalized D'Arcy and Watt (GDW) model. Each of them is studied using the fitting tools of the OriginLab Software. The fitting parameters are extracted with their errors while the accuracy of the fit is given by the R-squared coefficient.

*2.2.2. The Brunauer-Emmett-Teller (BET) model*

Originally developed to describe the physical adsorption of gas molecules on solid surfaces, the BET model can be adapted to seed biology in the context of moisture sorption isotherms. Seeds, like many other biological materials, have the ability to adsorb or desorb water from their environment depending on the surrounding relative humidity. This behavior can be graphically represented using moisture sorption isotherms. Just as the thermodynamical BET model considers the multilayer adsorption of gas molecules on solid surfaces, seeds can similarly adsorb water molecules in multiple layers. The first layer of water molecules can be directly bound to the seed's constituents (carbohydrates and proteins), while the subsequent layers might be due to weaker forces, such as hydrogen bonding or van der Waals forces between the water molecules themselves. In the context of seeds, the BET equation can be used to describe the relationship between the equilibrium moisture content of the seed WC %$_{DW}$ (instead of volume of gas adsorbed) and the relative humidity of the environment $H_{rel}$ (%) (instead of the pressure in the original equation). The resulting model is given by formula {1} [35]:

$$WC = \frac{M_m \times C_{BET} \times H_{rel}}{(1-H_{rel})(1+(C_{BET}-1)H_{rel})}$$

$$WC = \frac{M_m \times C_{BET} \times H_{rel}}{1+(C_{BET}-2)H_{rel}-(C_{BET}-1)H_{rel}^2} \quad \{1\}$$

$M_m$ is the monolayer moisture content and $C_{BET}$ is a dimensionless constant which relates to the energy of adsorption. The BET model is relevant to assess WC values for $H_{rel} < 50\%$ and is no longer accurate beyond [36].

### 2.2.3. The Guggenheim-Anderson-de Boer (GAB) model

The GAB model is a widely used mathematical model for describing the sorption isotherms of foodstuffs and agricultural products, including seeds. It is an extension of the BET model and is applied to moisture sorption data, especially at higher relative humidity levels. The GAB model is often preferred over the BET model for its better fit to experimental data across a wider range of relative humidity values. The GAB model considers that water sorption is achieved in a multilayer region, and is ruled by equation {2}:

$$WC = \frac{M_m \times C_{GAB} \times K \times H_{rel}}{(1 - K\,H_{rel})(1 - K\,H_{rel} + C_{GAB}\,K\,H_{rel})} \quad \{2\}$$

$M_m$ is the monolayer moisture content, as in the BET equation. It indicates the water content at which monolayer adsorption is complete. $C_{GAB}$ is a dimensionless constant, related to the enthalpy of sorption for the molecules in the multilayer, while K (dimensionless constant) represents the ratio of the sorption enthalpy of water in the multilayer to that in the monolayer. Usually, the GAB model is valid between 5 and 90 % of relative humidity [36].

### 2.2.4. D'Arcy-Watt (D&W) model

The D'Arcy-Watt model is employed to understand the water binding characteristics in seeds. It essentially aids in describing how water is sorbed (or absorbed) onto heterogeneous seed tissues, considering the varying energies that exist across multiple sorption sites [37]. Since seeds absorb moisture, they first bind water strongly at individual sites, then in weaker clusters, and finally as larger assemblages of water molecules. Understanding these regions and how seeds transition between them provides insights into how seeds respond to various hydration levels and conditions. The D'Arcy-Watt equation (see formula {3}) is composed of three separate terms. Each of these terms signifies a specific mode of water binding and has to be correlated with a region of the sorption isotherm:

- The 1st term designates the sites in seed tissue characterized by a high affinity for water. The more strongly individual water molecules bind to these sites, the less likely they are to be eliminated, making this type of binding important for the seed's resistance to desiccation or hydration. Parameters K and K' represent the affinity and total number of these high-affinity sites, respectively.
- The 2nd term corresponds to the sites with a lower affinity to water than those in the first region, but allowing the attachment of groups or clusters of water molecules to seed tissue. Parameter c represents both the number and strength of these weaker binding sites. Water bound in this way can be more easily removed or added, making it more dynamic.
- The 3rd term represents the final stages of hydration, where water not only binds to the seed, but condenses in significant quantities. This is macromolecular sorption, i.e. sorption in which water behaves more as a collective entity than as individual molecules. As a result, water condenses on the sites in the form of large assemblies/collections of molecules. Parameters k and k' indicate the affinity and number of sites respectively.

$$WC = \frac{KK'H_{rel}}{1+KH_{rel}} + cH_{rel} + \frac{kk'H_{rel}}{1-kH_{rel}} \quad \{3\}$$

$$N_{strong} = \frac{K'.N_A}{M} \quad \{4\}$$

$$N_{weak} = \frac{c.N_A}{M.p_s} \quad \{5\}$$

$$N_{multi} = \frac{k'.N_A}{M} \quad \{6\}$$

The numbers of strong sorption sites, weak sorption sites and multimolecular sites are given by equations {4}, {5} and {6} respectively, where $N_A$ is the Avogadro number, M the molecular weight of water per sorption site (in grams) and $p_s$ the vapour pressure at the studied temperature (2338.8 Pa at 20 °C). WC values can be accurately estimated on the full $H_{rel}$ range.

### 2.2.5. Generalized D'Arcy and Watt (GDW) model

The Generalized D'Arcy and Watt (GDW) model offers an evolved perspective on seed hydration dynamics compared with the traditional D&W model. It considers that seeds have Langmuir-type primary adsorption sites on their inner and outer surfaces. Each primary site has the particularity of being able to adsorb a single molecule of water. What is more, once a water molecule is bound to a primary site, it can in turn become a secondary adsorption site. The GDW model is based on two assumptions: (i) each water molecule adsorbed on a primary site does not necessarily form a subsequent secondary site, (ii) a single water molecule bound to a primary site can give rise to several secondary adsorption sites, resulting in a cascade hydration mechanism.

The GDW model is described by formula {7} as proposed by Furmaniak [38]:

$$WC = \frac{a_{mL}p_sK_LR_H}{1+p_sK_LR_H}\frac{1-c(1-\omega)R_H}{1-cR_H} \quad \{7\}$$

Where $a_{mL}$ (mol.g$^{-1}$) is the surface concentration of primary Langmuir-type adsorption sites, $K_L$ (Pa$^{-1}$) is the Langmuir constant, which signifies the strength of interaction between water and the seed tissue, c (dimensionless) is a kinetic constant related to the adsorption on the secondary sites and ω (dimensionless) designates the proportion of water molecules bound to primary sites that subsequently transform into secondary sites. One of the features of the GDW model, which sets it apart from its counterparts, is its global coverage of the sorption isotherm - from the driest state at 0% relative humidity to full saturation at 100%. Another significant advantage lies in its ability to deduce the primary site adsorption value, $a_{prim}$ (mol.g$^{-1}$), directly from $a_{mL}$ using the formula {8}:

$$a_{prim} = \frac{a_{mL} K_L p_s}{1 + K_L p_s} \quad \{8\}$$

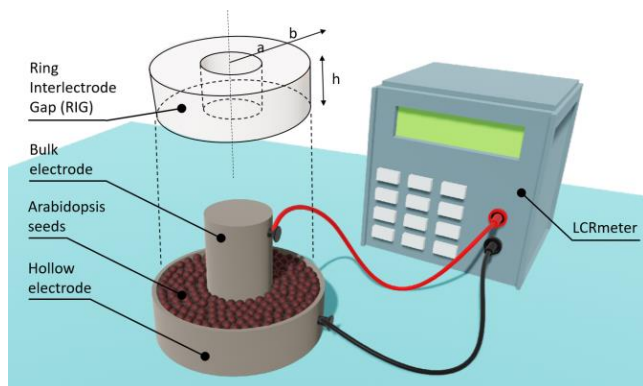

*Figure 1. Sketch of the coaxial cylindrical capacitor (CCC) containing Arabidopsis seeds in its interelectrode space to achieve their LCR measurements. (a = 2 mm, b = 5 mm, h = 3mm)*

## 2.3. LCR measurements

When using an LCRmeter, certain parameters such as resistance, capacitance and loss tangent are directly displayed while the others (complex permittivity, conductivity) require further calculations and derivations based on the direct measurements provided by the instrument [39]. For this reason, a coaxial cylindrical capacitor (CCC) has been designed following the model of Zhu *et al.*, as sketched in **Figure 1** [27]. The setup includes an inner bulk electrode (a = 2 mm, h = 3 mm), an outer hollow electrode (b = 5 mm, h = 3 mm) and a polymer assembly (polylactic acid) to nest them. The resulting ring interelectrode gap (RIG) has a volume of $\pi . (b^2 - a^2). h = 200\ mm^3$ where Arabidopsis seeds can be stacked. The two electrodes are connected to an LCRmeter (HM-8118 model from Rohde & Schwarz company) to measure seed electrical parameters over a 20 Hz-160 kHz range, in the parallel equivalent circuit mode. The measurement process is as follows:

- First, all frequencies are calibrated in open circuit (with the two clips apart) and subsequently in short circuit (with the two clips joined)
- Second, resistance and capacitance of the RIG are measured without seeds
- Third, resistance and capacitance of the RIG filled with 100 mg of seeds are measured

In this article, all the LCR measurements are achieved for three values of water contents: low water content at 3 %$_{DW}$, medium water content at 10 %$_{DW}$ and high water content at 30 %$_{DW}$. In each WC case, the LCR measurements are achieved in quadruplicate. For each replicate, the measurements are achieved a few hours before and a few hours after the plasma treatment.

## 2.4. Plasma source and electrical probes

A dielectric barrier device (DBD) in a plan-to-plan configuration with a gap of 1 mm is used in this work. As shown in **Figure 2**, it is composed of a 2mm thick dielectric barrier and two electrodes: an electrode biased to the high voltage (stainless-steel mesh, 30 × 40 cm$^2$ in area, with a 100 µm nominal opening, a wire diameter of 71 µm and an open area of 34 %) and a counter-electrode which is grounded (bulk alumina plate). The mesh electrode is powered with a high voltage generator composed of a function generator (ELC Annecy France, GF467AF) and a power amplifier (Crest Audio, 5500W, CC5500). The power supply is completed by a ballast resistor (250 kΩ, 600 W) to limit current.

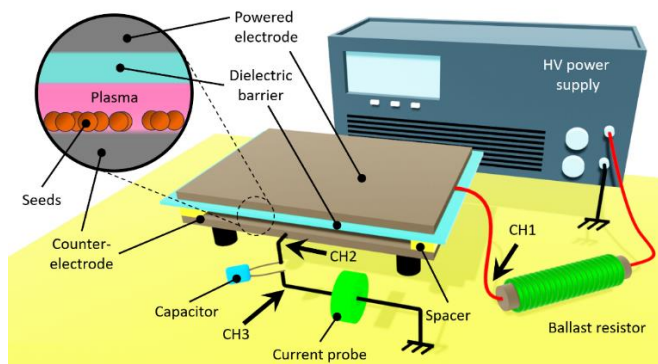

*Figure 2. Experimental setup of the dielectric barrier device (DBD) used to generate ambient air cold plasma and treat Arabidopsis seeds.*

Voltages and currents of the DBD are measured using a digital oscilloscope (model wavesurfer 3054 from Teledyne LeCroy company), with a 500 MHz bandpass and a 4GS/s sample rate. As sketched in **Figure 2**, three channels (CH) of the oscilloscope are utilized for the electrical characterizations. The voltages of the powered electrode and counter-electrode are measured using high voltage probes connected at CH1 and CH2 respectively (Tektronix P6015A 1000:1, Teledyne LeCroy PPE 20 kV 1000:1) while the discharge current is monitored at CH3 using a current transformer (Pearson company, model 2877) placed between the counter-electrode and the ground. A measurement capacitor (80 nF) is inserted between the counter-electrode and the ground to plot Lissajous curves. Their closed contour area is calculated to deduce the energy dissipated in the DBD. Multiplying this value by the operating frequency provides the electrical power injected in the DBD. The DBD source could process large quantities of Arabidopsis seeds, although each treatment is limited to 500 mg of Arabidopsis seeds (during 15 min) for the purposes of our experiments.

## 2.5. Statistical analysis

All the statistical analyses are performed using OriginLab Software statistical tools, including curve fitting and Mann-Whitney tests to determine significant differences on experimental data from control group and plasma group (water sorption isotherms).

# 3. Results & Discussion

## 3.1. Seed water content modifies the electrical parameters of the cold ambient air plasma (C2AP)

Arabidopsis seeds are treated by C2AP in the dielectric barrier device, sketched in **Figure 2**. Four scenarios are considered: DBD without seeds and DBD with 500 mg of Arabidopsis seeds either at low water content (3 %$_{DW}$), medium water content (10 %$_{DW}$) and high water content (30 %$_{DW}$). In each scenario, the seeds are spread all over the counter-electrode surface to guarantee an optimal interplay with plasma.

**Figures 3a** and **3b** illustrate the voltage and current profiles as a function of time in the DBD, with and without Arabidopsis seeds in the gap. $V_{DBD}$ typically varies between –7 and 7 kV with minor temporal variations in the four cases, considering a frequency of 150 Hz in all cases. The $I_{DBD}$ current profile comprises both a capacitive component ($I_{capa}$, characterized by a sinusoidal continuum) and a conductive component ($I_{cond}$, characterized by a series of very distinct peaks of the order of ten mA). These peaks provide information on the filamentary nature of C2AP (**Figure 3b**). During each period, these peaks appear in two temporally well-defined regions known as "active phases" ($\tau_a$ regions in **Figure 3b**). Conversely, "quenching phases" feature the absence of current peaks ($\tau_q$ regions in **Figure 3b**). The main findings in **Figure 3b** are as follows: (i) in the absence of seeds, positive and negative current peaks are of comparable intensity, (ii) with seeds, positive current peaks are more intense than their negative counterparts, and (iii) a tenfold increase in seed hydration (from 3 %$_{DW}$ to 30 %$_{DW}$) doubles peak amplitude.

To take this a step further, the current peaks in the four previous scenarios are counted over 10 periods. This counting is achieved by processing the gross data of **Figure 3b** with Origin software. More specifically, two analytical methods are combined: asymmetric least squares smoothing baseline (asymmetric factor = 0.2, threshold = 0.05, smoothing factor = 9, number of iterations = 8) and peaks finding (method: 2nd derivative, Smoothing window size = 0, Direction = Both, Method filtering = By height, threshold height = 20%). A statistical representation of these current peaks is proposed in **Figure 3c** under the form of data-violin plots. The width of the distributions (i.e. violins) at different values indicates the density of the data at that value, with wider sections representing higher density (more data points). In the absence of seeds, the distribution of current peaks ranges from –20 mA to +20 mA, with a propensity for the positive current peaks. Introducing dry seeds (3 %$_{DW}$) in the gap enhances the asymmetric distribution which ranges from –15 mA to +30 mA with an average value of 9 mA. Increasing water content to 10 %$_{DW}$ tightens the peaks between –10 mA and +25 mA with a larger number of peaks at values close to 10 mA. At 30 %$_{DW}$, current peaks are concentrated in the same interval but with an ever increasing number of positive current peaks at 7 mA and an average at 5 mA. It is also interesting to note that, according to **Figure 3b**, this peak distribution covers a wider active phase compared with dry seeds at 3%$_{DW}$.

The capacitive component of the current is shown in **Figure 3d**. It is proportional to the derivative of the DBD voltage ($I_{capa}$ = C.dV$_{DBD}$/dt) and extracted from $I_{DBD}$(t) following the "asymmetric least squares smoothing baseline" method. $I_{capa}$ shows the higher magnitudes in presence of seeds, especially when their hydration reaches 30 %$_{DW}$. As further validated in the LCR measurements (Section 3.3), increasing seed water content rises the capacitance of the seeds in addition to that of the dielectric barrier. This explains why at 30 %$_{DW}$, the capacitive component of current has a magnitude between –8 mA and 10 mA. Interestingly, the temporal profiles of these capacitive currents present a quasi-sinusoidal appearance, i.e. a periodic but slightly asymmetrical profile, particularly in the "no seeds" case. A first hypothesis could be that after each half-period of the applied voltage, the accumulation of charges on the dielectric barrier alters the capacitive current in the subsequent half-period. However, this assumption is poorly convincing because, as detailed hereafter in **Figure 3e**, the Lissajous curves are precisely centered on the (0,0) origin. A more relevant hypothesis to explain asymmetry is secondary processes such as ion-induced secondary electron emission that dominates if no seeds are introduced in the gap. When hitting the dielectric surface, the energetic ions are likely to release secondary electrons that participate in the discharge. The presence of seeds, especially at higher WC, mitigates this effect, by increasing the overall value of the DBD capacitance.

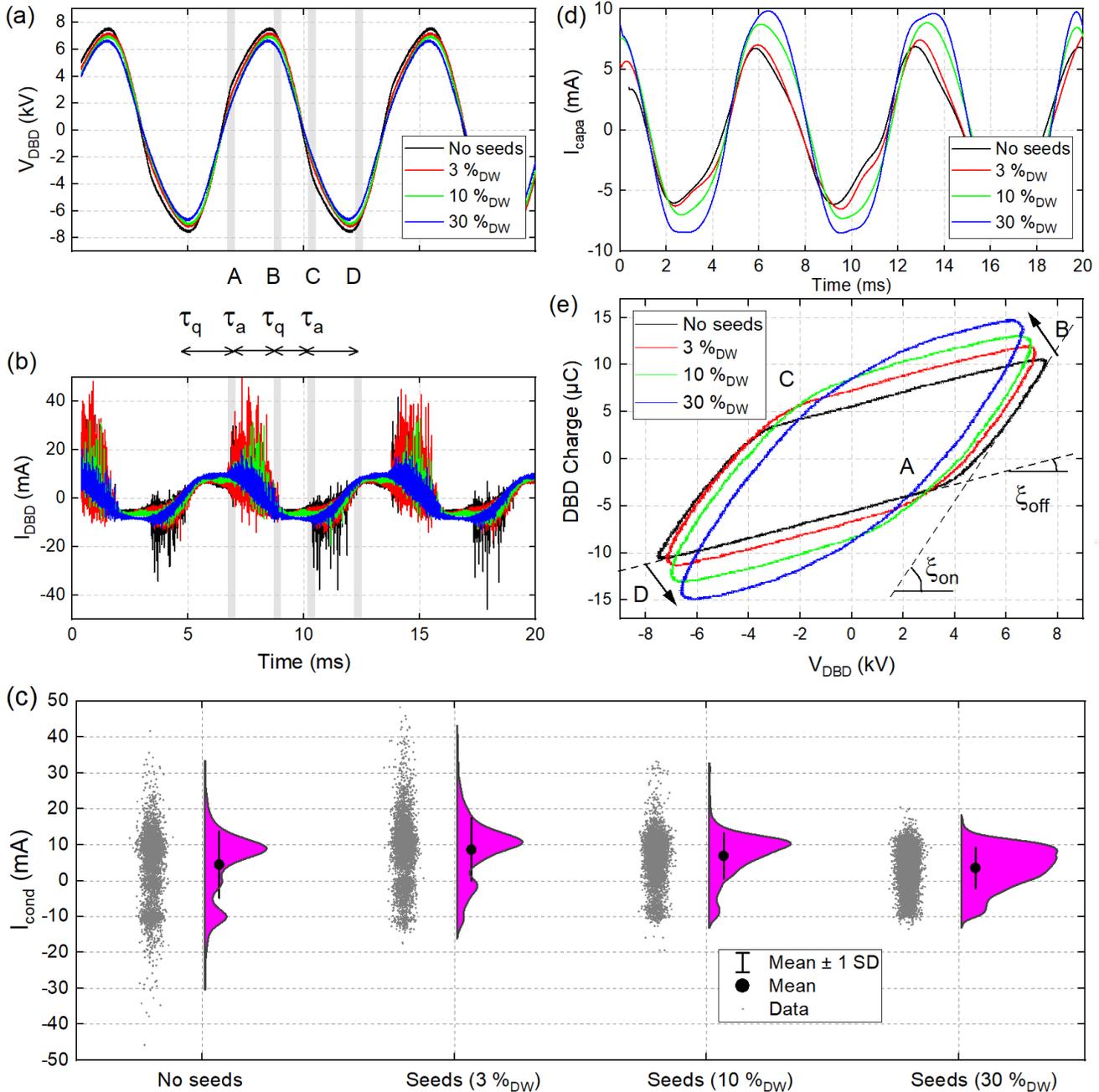

Figure 3. (a-e) Electrical characterizations considering four scenarios: no seeds in the gap, seeds at 3 %$_{DW}$, seeds at 10 %$_{DW}$, seeds at 30 %$_{DW}$. (a) Temporal profile of DBD voltage at 150 Hz, (b) Temporal profile of DBD current at 150 Hz, corresponding to the superimposition of a capacitive component (sine signal) and a conductive component (peaks). (c) Distribution and semi-violon-plots of the current peaks ($I_{Cond}$: conductive component of $I_{DBD}$) in the four previous scenarios , (d) Temporal profile of $I_{DBD}$ capacitive component, (e) Lissajous curves of the DBD.

The measurement capacitor ($C_m$) depicted in **Figure 2** is utilized to plot the electrical charge of C2AP as a function of $V_{DBD}$, as shown in **Figure 3e**. For each of the four scenarios, a closed contour is obtained, also known as the Lissajous curve, whose area multiplied by the voltage frequency corresponds to the electrical power deposited in the DBD. To analyze these Lissajous curves, we first need to define the main elements of the DBD. The volume of the gap ($V_{gap}$) is defined as the volume of gas ($V_{gas}$) with/without the volume of seeds ($V_{seeds}$), following equation {9}. The gas volume ($V_{gas}$) corresponds to the volume of air in the gap, including the regions that are ionized ($V_{pl,on}$) and eventually non-ionized ($V_{pl,off}$), as suggested in equation {10}. Therefore, when seeds are present in the gap, the fraction of non-ionized gas is given by equation {11}. The capacitance of the dielectric barrier ($C_{diel}$) is given by equation {12} where $\varepsilon_0$ is the vacuum permittivity (8.854 × 10$^{-12}$ F.m$^{-1}$), $\varepsilon_r$ the relative permittivity of glass (6), A the electrode area (30 × 40

cm$^2$) and $t_{diel}$ the thickness of the dielectric barrier (1 mm). Depending on whether or not it contains seeds, $C_{gap}$ is given by equation {13} which corresponds to the paralleling of the capacitance of the confined gas ($C_{gas}$) and the capacitance of the seeds ($C_{seeds}$). $C_{gas}$ and $C_{seeds}$ are modulated by the volumes they respectively occupy within the gap, i.e. $V_{gas}/V_{gap}$ and $V_{seeds}/V_{gap}$ respectively. For more information, comprehensive insights are available in the equivalent electrical model of the seed-stacked DBD [20].

In **Figure 3e**, each Lissajous curve is characterized by :
- A consistent power output of 20 W
- A closed contour shape very close to a parallelogram in the absence of seeds, transitioning to an oval-like shape for seeds showing increased hydration levels.
- Edges [BC] and [AD] that are associated with the quenching phase in **Figure 3b** and whose slope is equal to the capacitance of the DBD in the absence of plasma ($\xi_{off}$). As written in equation {14}, this capacitance aligns with the capacitance of the dielectric barrier ($C_{diel}$) in series with the gap capacitance ($C_{gap}$). Graphically, its value is estimated to 718 pF.
- Edges [AB] and [CD] that are associated with the active phase in **Figure 3b** and whose slope is equal to the capacitance of the DBD when the plasma is ignited ($\xi_{on}$), i.e. the capacitance of the dielectric barrier in series with a fraction of the gap capacitance, as underlined in equation {15}. Graphically, its value is estimated to 2250 pF. The capacitance of the seeds ($C_{seeds}$) can be roughly estimated to 6 nF considering equation {16} that is directly inferred from equation {15} [40]. A more accurate estimation of $C_{seeds}$ will be achieved by LCRmetry in Section 3.3.

$$V_{gap} = V_{gas} + V_{seeds} \quad \{9\}$$

$$V_{gas} = V_{pl,on} + V_{pl,off} \quad \{10\}$$

$$\eta = \frac{V_{gas}}{V_{gap}} \cdot \frac{V_{pl,off}}{V_{gap}} \quad \{11\}$$

$$C_{diel} = \varepsilon_0 \cdot \varepsilon_r \cdot \frac{A}{t_{diel}} \quad \{12\}$$

$$C_{gap} = \left(\frac{V_{gas}}{V_{gap}}\right) \cdot C_{gas} + \left(\frac{V_{seeds}}{V_{gap}}\right) \cdot C_{seeds} \quad \{13\}$$

$$\frac{1}{\xi_{off}} = \frac{1}{C_{diel}} + \frac{1}{C_{gap}} \quad \{14\}$$

$$\frac{1}{\xi_{on}} = \frac{1}{C_{diel}} + \frac{1}{\eta \cdot C_{gas} // C_{seeds}} = \frac{1}{C_{diel}} + \frac{\eta \cdot C_{gas} + C_{seeds}}{\eta \cdot C_{gas} \cdot C_{seeds}} \quad \{15\}$$

$$C_{seeds} = \frac{1}{\frac{1}{\xi_{on}} - \frac{1}{C_{diel}} - \frac{1}{\eta \cdot C_{gas}}} \quad \{16\}$$

For any given Lissajous curve in **Figure 3e**, the point A is determined by the intersection of the line with the $\xi_{off}$ slope and the one with the $\xi_{on}$ slope. This boundary becomes less distinct as seed water content is increased, indicating:
- The value of $\xi_{on}$ can change along the active phase ($\tau_a$). In absence of seeds, the streamers directly bridge the dielectric barrier and the grounded counter-electrode. However, this bridging becomes no longer satisfactory in the presence of seeds, especially if they are moist. It can be argued that the 30 %$_{DW}$ seeds have a very low electrical resistivity, placing them at a potential close to the ground. Consequently, at the onset of each active phase, the initial streamers may preferentially form between the dielectric barrier and the top of the seeds, instead of bridging the full gap thickness. Then, as the $V_{DBD}$ voltage increases during the active phase, streamers progressively cover longer distances, eventually linking the dielectric barrier to the grounded counter-electrode. This explains the time-varying profile of $\xi_{on}$ during the active phase.
- Seeds at 30 %$_{DW}$ allow streamers to be created at smaller $V_{DBD}$ values than in the other three scenarios. This last observation is in line with the result in **Figure 3b**, where the active phase ($\tau_a$) increases sharply at the expense of the quenching phase.

To summarize, seed hydration has a significant impact on electrical profiles, streamers formation and overall DBD performance. Increased WC values promote the capacitive component of IDBD while filtering out the generation of current peaks higher than 10 mA and maintaining a plasma power output of 20 W. Our results also indicate that increasing WC enhances the ability of seeds to attract the streamers. At 30%$_{DW}$, there is an increase in the interaction of low-energy positive streamers with the seeds themselves, rather than these streamers primarily forming connections between the dielectric barrier and the counter-electrode. This means that at this moisture level, the plasma treatment more directly affects the seeds, potentially enhancing the treatment's effectiveness in improving seed germination and breaking seed dormancy, as further evidenced in the following section.

## 3.2. Arabidopsis seed dormancy is released by plasma, especially in seeds with high water content

C2AP treatments (7 kV in amplitude and 150 Hz in frequency) are applied for 15 minutes on the dormant Arabidopsis seeds, categorized into the three previously mentioned groups (3 %$_{DW}$, 10 %$_{DW}$ and 30 %$_{DW}$). **Figure 4** reports the germination monitoring for these groups upon a period of 12 days. Whether for control or plasma groups, the germination rate is consistently increased, with a more pronounced effect at higher water content values. Hence, in the case of seeds at 30 %$_{DW}$, the **Figure 4** shows that plasma treatment increases the germination rate from 22 % to more than 65 % at day 10.

In a prior study dedicated to Arabidopsis seeds, we have established that increasing WC make the seeds enter into an amorphous state, thereby reducing their cytoplasmic viscosity [10]. This so-called "rubbery state" make the seeds more receptive to the effects of C2AP, especially to release their dormancy. In addition to this effect, the Section 3.1. of this article clearly shows that higher WC reduces the overall intensity of the current peaks but significantly increases their number, directing them towards the seeds. Therefore, higher seeds moisture can also enhance the electrical properties of C2AP and therefore its efficiency in releasing seed dormancy.

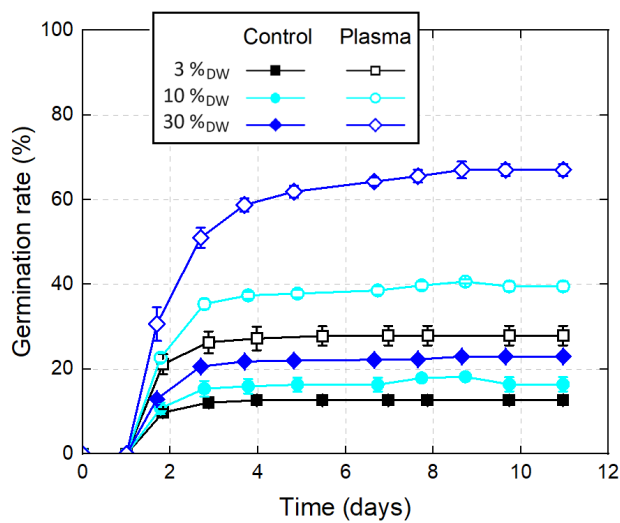

*Figure 4. Germination curves of dormant Arabidopsis seeds with different water contents (3 %$_{DW}$, 10 %$_{DW}$, 30 %$_{DW}$), whether untreated (Control) or exposed to cold ambient air plasma for 15 min (Plasma). Plasma conditions: 7 kV in amplitude, 150 Hz in frequency.*

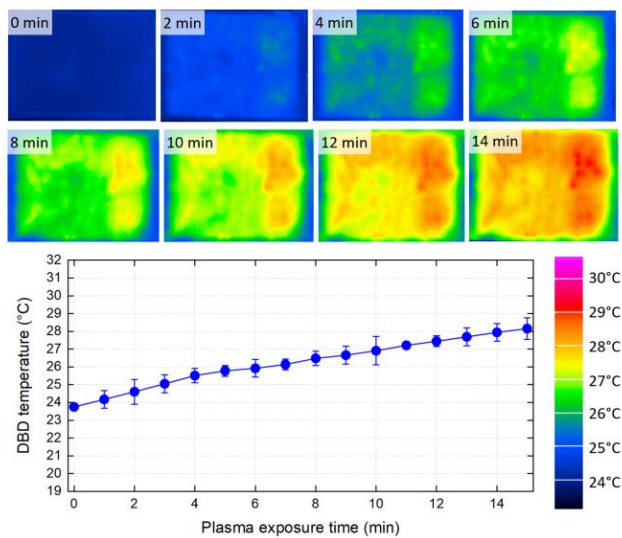

*Figure 5. Variation of the DBD temperature (powered electrode temperature) as a function of the plasma exposure time. C2AP at 7 kV of amplitude, 150 Hz.*

Additionally, it is important to emphasize that, irrespective of the moisture content being 3%$_{DW}$, 10%$_{DW}$, or 30%$_{DW}$, the seeds undergo treatment at a plasma gas temperature that remains very close to the ambient air temperature. This is demonstrated in **Figure 5**, where infrared imaging was used to track the temperature changes of the powered electrode over a 15-minute period under the same scenarios presented in **Figures 3** and **Figure 4**. The data shows a modest temperature increase from 23.75 °C to just 28.15 °C following a 15-minute exposure to plasma, thereby suggesting that the temperature reached does not rise to levels detrimental to the seeds.

## 3.3. LCRmetry explains how seed bioelectrical parameters change plasma, but not vice versa

In Section 3.1., we have shown that WC modifies the plasma electrical properties (streamers distribution), probably because it changes seeds resistance and capacitance. To verify this hypothesis, LCR measurements are performed directly on dormant Arabidopsis in the RIG device sketched in **Figure 1** and following the procedure detailed in Section 2.3. These measurements were taken on seeds a few hours before and a few hours after exposure to C2AP, and may therefore not be extrapolated to larger time scales. For remind, whether considered one by one or as a batch, seeds can be electrically modeled by a resistor in parallel with a capacitor [20]. The following dielectric parameters are related to the bulk properties of seeds (by opposition to their external surface properties) and have been measured in response to an external electric field: resistance (R), capacitance (C), complex relative permittivity ($\varepsilon_r$), tangent loss (tan δ) and conductivity (σ).

### 3.3.1. Resistance and capacitance

The resistance ($R_{seeds}$) and capacitance ($C_{seeds}$) of the Arabidopsis seeds are plotted versus frequency in **Figure 6a** and **5b** respectively. At 150 Hz, which corresponds to the frequency of the plasma's high voltage (Section 3.2.), LCRmetry indicates resistance values of 1050 Ω, 607 Ω, 20 MΩ and capacitance values of 3.3 nF, 5.2 nF, 46.4 nF for moisture contents of 3 %$_{DW}$, 10 %$_{DW}$ and 30 %$_{DW}$ respectively. In the specific case of $C_{seeds}$, LCRmetry provides therefore a more accurate value than the one estimated to roughly 6 nF by equation {16}.

Whether the seeds are exposed or not to plasma, they show no significant difference in their resistance and capacitance values over the 20 Hz-150 kHz range. This observation warrants caution as one might deduce that plasma does not modify the bulk properties of seeds, only affecting the seeds outer layers (coating and pericarp). Overall, this idea is sustained by synchrotron radiation X-ray tomographic microscopy that demonstrates no significant difference between seeds untreated and treated by plasma, whether for broccoli or wheat [41], [42]. However, a fine analysis achieved on Arabidopsis demonstrates that plasma can actually enlarge the intercellular spaces, driving to a slight increase in porosity [43]. So, even if LCRmetry indicates no discrepancy between control and plasma groups (in terms of $R_{seeds}$ and $C_{seeds}$ variations), bulk properties may still be changed as long as these modifications do not bring or remove significant amounts of new materials. More specifically and considering that seeds are tridimensional porous biomaterials, C2AP is likely to chemically modify both inner and outer seed tissues by attaching oxygenated groups (derived from the plasma's reactive oxygen species), although such changes are subtle enough to elude LCRmetry detection, especially at 3 %$_{DW}$ and 10 %$_{DW}$. Owing to the overlap between the control and plasma curves (see **Figure 6a** and **6b**), one can reasonably admit that plasma does not significantly change the relative contents in starch, proteins and lipids, at least in the hours following the plasma treatment. However, it is worth stressing that discrepancies could arise if R$_{seeds}$ and C$_{seeds}$ were measured a few days or a few weeks after plasma treatment.

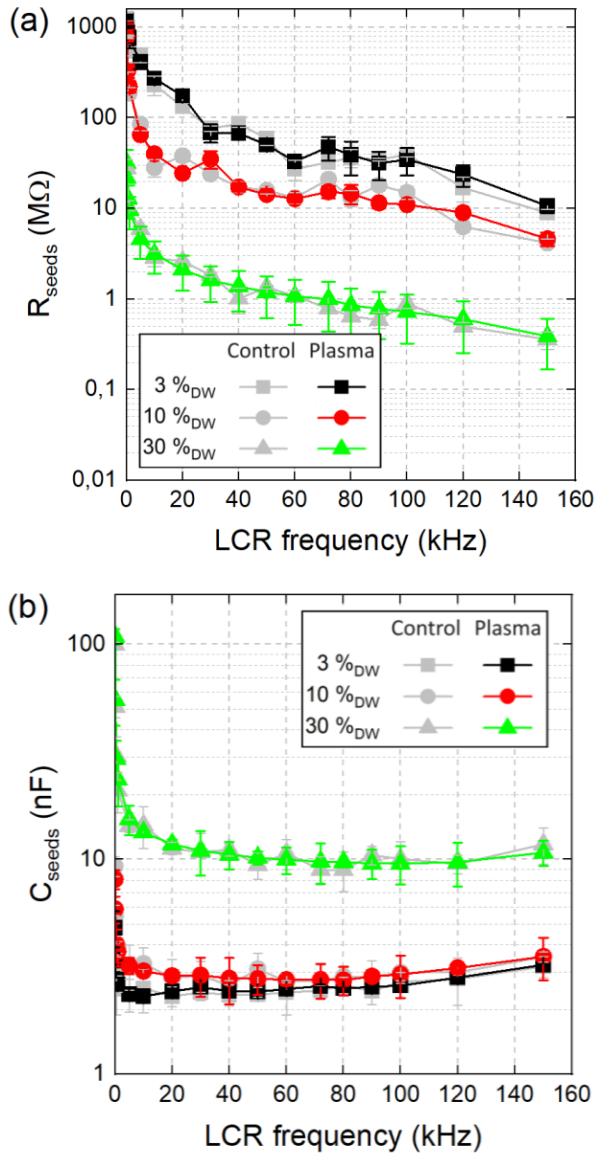

*Figure 6. Variation of (a) seeds resistance and (b) seeds capacitance as a function of LCR frequency for six distinct groups categorized by their water content levels (3 %$_{DW}$, 10 %$_{DW}$, 30 %$_{DW}$) and their exposure (or not) to plasma. Means of 4 replicates, each consisting of 100 mg of Arabidopsis seeds ± SD.*

At lower frequencies, the charge carriers in the seeds have more time to align with the electric field, because its direction changes less frequently. Hence the dipoles (polar molecules) within the seed material have more time to orient themselves in response to the electric field. As a result, the seed exhibits higher polarization and, consequently, higher R$_{seed}$ and C$_{seed}$ values. Whether for control or plasma-treated seeds, as the LCR frequency is increased, this alignment becomes less effective, leading to a decrease in both R$_{seed}$ (**Figure 6a**) and C$_{seed}$ (**Figure 6b**). For seeds at 30 %$_{DW}$, the resistance decays from 50 MΩ to 400 kΩ while the capacitance drops significantly from approximately 100 nF (20 Hz) to 15 nF (1 kHz). These trends could result from the polar nature of water and its interactions with cellular structures, implying a vanishing of the polarization mechanisms such as membrane polarization. In DC (not discussed in this article), the polarization effects could make the capacitive components of the seeds behave as open switches so that current flow only propagates through the purely resistive components. At low frequencies, in the 20 Hz-1 kHz range, these capacitive components dominate, and behave as very high resistive components which therefore impede current flow. At higher frequencies, starting from a few 10 kHz in **Figure 6a** and 6b, the capacitive components show a weaker impedance and can be modeled as low-resistive components, hence allowing current to flow more easily. At very high frequencies, typically in the GHz range (not discussed in this article), these capacitive components could ultimately behave as closed switches, hence short-circuiting R$_{seeds}$. The Section 3.3.5 provides a comprehensive schematics of the seed equivalent electrical model (**Figure 9**) and how its capacitive component behaves depending on frequency (DC, low, high, very high frequency).

At a fixed value of LCR frequency, increasing seed water content leads to lower resistance and higher capacitance. The reason is that the seeds contain more free water which facilitates ionic conduction (ion movement), whether for the control or the plasma-treated seeds. Seeds with 3 %$_{DW}$ have higher resistance while at 10 %$_{DW}$ and 30 %$_{DW}$, the resistance drops because of more conductive pathways. Capacitance rises with WC because water has a high dielectric constant. At 3 %$_{DW}$, the seeds have lower capacitance but for higher WC values, the seed's electrical charge storage capacity grows with the increasing water volume.

### *3.3.2. Complex relative permittivity ($ε_r$), dielectric constant ($ε_r'$), dielectric loss factor ($ε_r''$)*

The complex relative permittivity is a crucial parameter for understanding seed dielectric and conductive properties under an electric field. It consists of the dielectric constant ($ε_r'$) and the dielectric loss factor ($ε_r''$), as expressed in formula {17} [44]. The ratio between these, termed the loss tangent (tan δ) from equation {18} [45], relates to the seed's electric power dissipation.

The dielectric constant ($ε_r'$) reflects the seed's capacity to store electrical energy, showing how polar molecules within seeds can adjust to an external electric field. According to equation {19}, $ε_r'$ is determined by the ratio of the capacitance of the RIG device with seeds (C) to the capacitance of the RIG device without seeds (C$_0$), all multiplied by the vacuum permittivity $ε_0$ = 8.854 × 10$^{-12}$ F.m$^{-1}$ [46]. **Figure 7a** indicates that at 30 %$_{DW}$, ε$_r$' strongly decreases on the 20 Hz - 10 kHz range and then slightly decays. This change is a key characteristic of dielectric relaxation, i.e. the time required for the polar entities within the seeds to respond to time variations of the external electric field. Above 20 kHz, most polarization mechanisms reach their limits, with only fast mechanisms such as electronic polarization, that are expected to occur. The seeds at 3 %$_{DW}$ and 10 %$_{DW}$ exhibit relatively flat and low profiles across the entire frequency range, suggesting that their dielectric response is less sensitive to frequency changes or that the concentration of the active dielectric component is less.

The dielectric loss factor (ε$_r$'') corresponds to the imaginary part of the complex relative permittivity and reflects the seed's propensity to dissipate the stored electrical energy when it is subjected to an alternating electric field [47]. This factor is crucial for understanding how the seeds' dielectric properties, influenced by their water content, interact with the plasma treatment. ε$_r$'' is derived from the ratio of two terms: (i) a numerator corresponding to the difference of the conductance with seeds (G) and without seeds (G$_0$), (ii) a denominator corresponding to the product of the angular frequency by the capacitance of RIG device without seeds (C$_0$) [48], as written in equation {20}. The **Figure 7** clearly shows that for the seeds at 30 %$_{DW}$, the stored electrical energy is more dissipated at lower frequency (< 10 kHz), essentially because the capacitive components of the seeds dominate. Beyond 10 kHz, the values of ε$_r$'' slightly decay while being maintained at high value (roughly 10$^{-3}$), which is consistent with **Figure 7a**. At a given LCR frequency, ε$_r$'' increases when rising WC, with values close to 6 × 10$^{-2}$ (at 30 %$_{DW}$) and 1 × 10$^{-4}$ (at 3 %$_{DW}$ and 10 %$_{DW}$) at 150 Hz. As detailed in the next section, this level of energetic dissipation can be attributed to both dielectric and conductive losses.

$$\varepsilon_r = \varepsilon_r' - j\varepsilon_r'' \quad \{17\}$$

$$\tan\delta = \frac{\varepsilon_r''}{\varepsilon_r'} = \frac{G - G_0}{2\pi f C} \quad \{18\}$$

$$\varepsilon_r' = \frac{1}{\varepsilon_0} \cdot \frac{C}{C_0} \quad \{19\}$$

$$\varepsilon_r'' = \frac{1}{\varepsilon_0} \frac{G - G_0}{2\pi f C_0} \quad \{20\}$$

$$\sigma = 2\pi f \varepsilon_0^2 \varepsilon_r'' \quad \{21\}$$

### 3.3.3. Dielectric/conductive losses & electrical conductivity (σ)

Electrical losses in seeds manifest through two primary mechanisms: dielectric and conductivity losses:
- The dielectric losses result from the time lag in the response of the seed's polar molecules to the applied electric field. Different relaxation processes can explain these dielectric losses, some of them occurring at high frequency and others at low frequency, as discussed hereafter.
- The conduction losses correspond to the energy lost when free charges, such as ions, move through the seed's internal structures. These losses are particularly evident in seeds with a high water content, which introduces more free ions. Under the influence of the external electric field, these charges move inside the seed, resulting in energy dissipation.

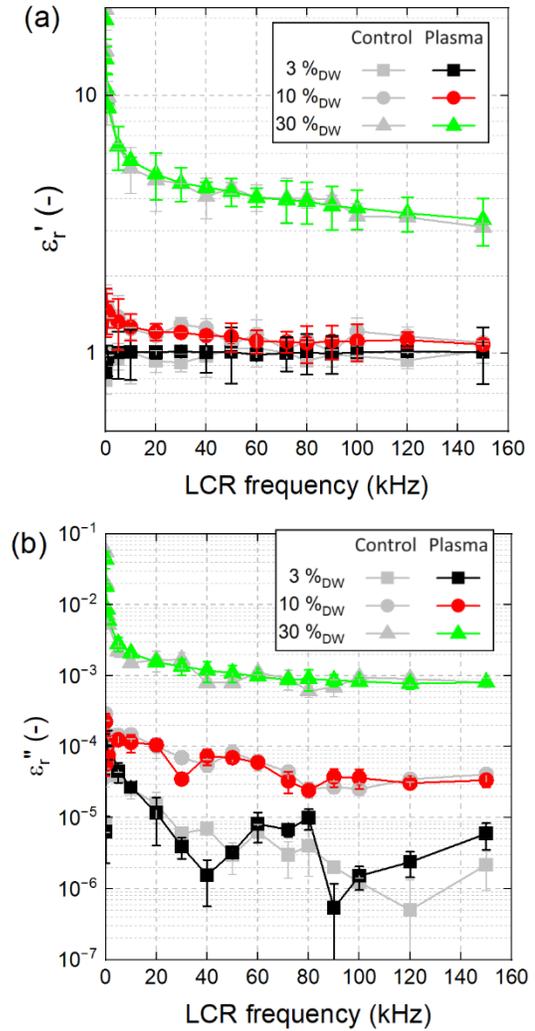

*Figure 7. Variation of (a) dielectric constant ε' and (b) dielectric loss factor ε'' as a function of LCR frequency for six distinct groups categorized by their water content levels (3 %$_{DW}$, 10 %$_{DW}$, 30 %$_{DW}$) and their exposure (or not) to plasma. Means of 4 replicates, each consisting of 100 mg of Arabidopsis seeds ± SD.*

Dielectric losses and conductive losses can be differentiated considering the variations of loss tangent (tan δ, in equation {18}) and electrical conductivity (σ, in equation {21}) over the LCR frequency range. A peak in tan δ at a particular frequency followed by a drop off indicates dielectric losses related to a relaxation process [49]. However, if tan δ is nearly constant and independent of frequency, it is indicative of conductive losses. Electrical conductivity (σ) provides a direct measure of how easily charges can move within the seed. At low frequencies or DC, a higher electrical conductivity indicates that the seed has more free charge carriers, leading to increased conductive losses [50]. At higher frequencies, if the conductivity remains relatively stable or increases linearly, then it indicates that conductive losses are dominant.

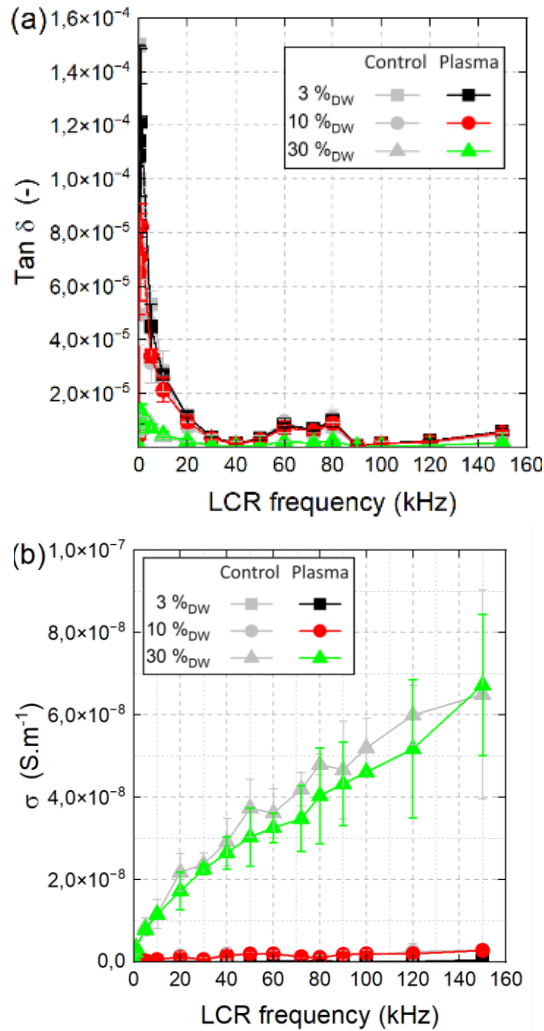

*Figure 8. Variation of (a) Loss tangent tan δ and (b) electrical conductivity σ as a function of LCR frequency for six distinct groups categorized by their water content levels (3 %$_{DW}$, 10 %$_{DW}$, 30 %$_{DW}$) and their exposure (or not) to plasma. Means of 4 replicates, each consisting of 100 mg of Arabidopsis seeds ± SD.*

In **Figure 8a**, the variations of tan δ remain very low over the full frequency range, always lower than $2.0 \times 10^{-4}$, hence suggesting that seeds are much more efficient at storing electrical energy than at dissipating it. The values of tan δ would have been much more elevated if the seeds were exposed to high-frequency electric fields, specifically in the microwave region (e.g. 2.45 GHz). For these very high frequencies, the water molecules struggle to reorient themselves, resulting in a kind of friction among them, i.e. a resistance to change in orientation that leads to energy being lost in the form of heat. Since microwave frequencies are neither reached upon the plasma treatments nor investigated as part of these LCR measurements, the values of tan δ remain very close to 0.

Still, **Figure 8a** shows values of tan δ close to $2.0 \times 10^{-4}$ for frequency lower than 10 kHz and therefore at 150 Hz which corresponds to the frequency at which plasma treatment is carried out. Even if a value of $2.0 \times 10^{-4}$ is low, it remains significantly higher than the detection threshold of the LCRmeter and therefore deserves investigation. At 150 Hz, the electric field changes slowly, allowing more time for dipoles to undergo processes like polarization relaxation. As this relaxation process dissipates energy (due to frictional effects as the dipoles move), it can contribute to an increase in the dielectric losses. This phenomenon is even more evident in drier seeds, because they have fewer water molecules, which need more time to reorient in response to the electric field. Several assumptions can be advanced to explain these low-frequency dielectric losses:

- At low frequencies, the slow oscillation of the electric field allows the water molecules to undergo complete orientation polarization, meaning they can rotate over large angle values to achieve full realignment with each period [49]. In the context of seed biology, these full realignments of water molecules occur within a cellular organic matrix, potentially leading to frictional losses, and thus to heat dissipation. The **Figure 9** indicates how this complete realignment can operate at low frequencies (Scenario A → B) by opposition to partial realignment at high frequencies (Scenario C → D).
- Seeds consist of cells with a lipid bilayer membrane. When an electric field is applied, the cell membrane itself can get polarized [51], [52]. At low frequencies, this polarization can lead to a specific type of relaxation called "Maxwell-Wagner-Sillars (MWS) polarization" [53]. This phenomenon may occur due to the accumulation of charges at the interfaces between the cell membrane and the cytoplasm or the cell wall, causing a delay in the alignment of dipoles, and therefore, energy dissipation.
- Within the seed cells, there are various ions, including those from dissolved nutrients and electrolytes. At low frequencies, these ions have more time to move in response to the applied electric field, leading to a relaxation mechanism. The movement of these ions within the cell and in intercellular spaces can cause frictional losses as they pass through the viscous cytoplasm or cell wall components.
- Seeds contain various macromolecules, especially proteins, starch, and nucleic acids. Some of these molecules, especially proteins, can undergo conformational changes or minor movements in response to an electric field [54]. Furthermore, water can be bound to these macromolecules, hence a restricted mobility and a response to electric field that is different from that of free water. At low frequencies, there is a greater possibility for these macromolecules to participate in relaxation processes, leading to energy dissipation.

The **figure 8b** shows a broad range of conductivity values, from $10^{-12}$ to $10^{-7}$ S/m. The seeds at 30 %$_{DW}$ indicates that electrical conductivity almost linearly increases with the LCR frequency. Such profile is characteristic of conductivity losses, i.e. the dissipation of energy through the displacement of charge carriers within the seed's internal structures. At 10 %$_{DW}$, the absence of free water within seed tissues results in $\sigma$ values close to $10^{-12}$ S.m$^{-1}$, which makes the conductivity losses negligible. This is even more true at 3 %$_{DW}$. In this article, the Arabidopsis seeds have been exposed to plasma supplied in high voltage at 150 Hz. At this frequency, the conductivity losses are negligible, whether at low, medium or high water content.

*3.3.4. Summary of the LCR measurements*

At low or medium water contents (3 %$_{DW}$, 10 %$_{DW}$), C2AP does not change the seeds dielectric parameters, especially their resistance and capacitance. This means that the relative contents in starch, proteins and lipids are not altered and that C2AP does not bring/remove new materials. However, it can modify surfaces, especially through chemical functionalization of both inner and outer seed tissues by attaching oxygenated groups (derived from the plasma's reactive oxygen species). The seed's capacity to store ($\varepsilon_r'$) and dissipate electrical energy ($\varepsilon_r''$) is enhanced at frequencies lower than 1 kHz, especially for the seeds at 30 %$_{DW}$. In this region, the seeds exhibit low but significant dielectric losses. These ones result from processes such as polarization relaxation where the polar molecules (essentially water) can completely realign with the slow electric field, thereby driving to frictional interactions with biomolecular structures (cellular organic matrix, lipid bilayer membrane, macromolecules and dissolved nutrients and electrolytes). Overall, the conductive losses remain very low below 1 kHz and therefore in the treatment conditions of Arabiodpsis seeds. The **Figure 9** proposes an overview of how Arabidospsis' equivalent electrical model behaves depending on the frequency of the applied high voltage. Especially, it permits to clearly decipher the dielectric losses operating at high frequencies (the most known) from those at low frequencies.

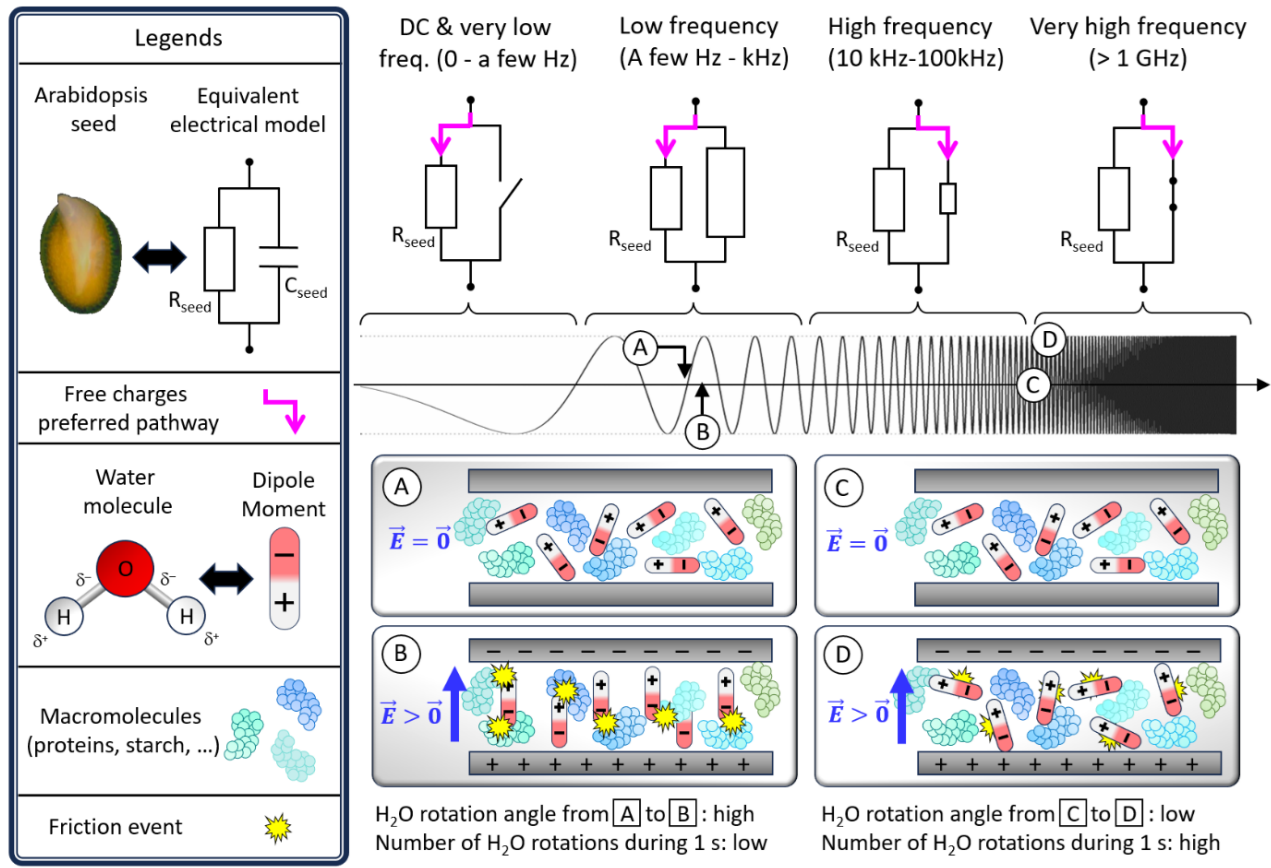

*Figure 9. Equivalent electrical model of Arabidopsis seeds and response of this model to different frequencies of the applied electric field. Detailed insets further depict how water molecules within the seed interact under different electrical conditions, emphasizing the rotation and frictional effects at cellular levels.*

## 3.4. Cold plasma from ambient air significantly changes seed water content only at low and high relative humidities

Another approach to probing the influence of C2AP on seed bulk properties, and in particular on the water content of internal tissues, is to perform sorption isotherms. Focusing on these sorption properties should help us to understand how WC can influence the plasma process efficiency in releasing the dormancy of Arabidopsis seeds. For this, seed water content is plotted as a function of relative humidity, following the protocol detailed in Section 2.2. As shown in **Figure 10a** (or **10b**, **10c**, **10d**), nine experimental conditions are studied considering untreated seeds (Control group) as well as plasma-treated seeds (Plasma group, 15 min of C2AP). These experimental datapoints are connected the ones to the others, using four sorption fitting models: the BET model (**Figure 10a**), GAB model (**Figure 10b**), D&W model (**Figure 10c**) and GDW model (**Figure 10d**).

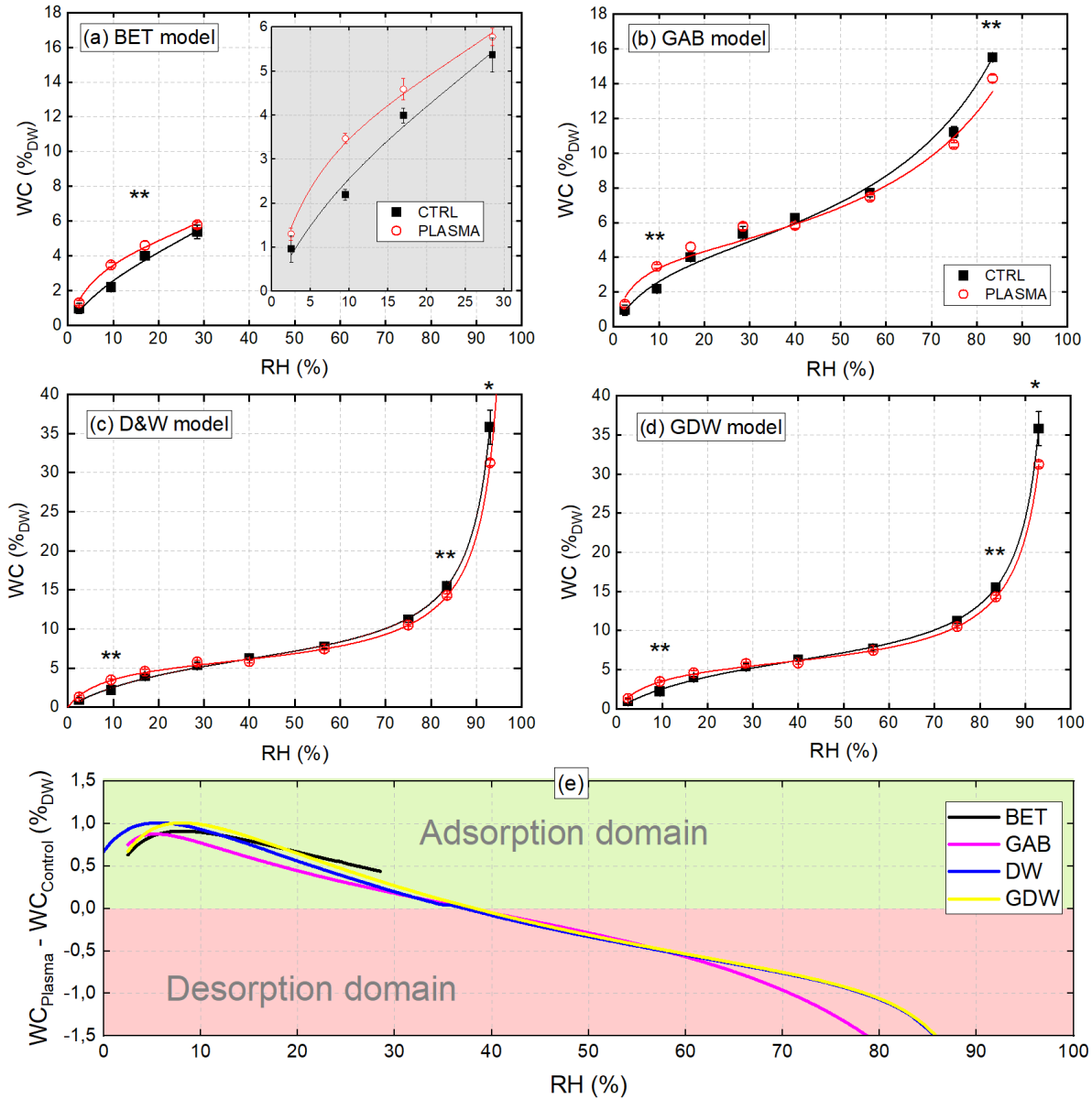

*Figure 10. (a-d) Water sorption isotherms plotted from experimental data points of dormant Arabidopsis seeds either non-treated (CTRL) or plasma-treated (PLASMA) at 20 °C. The water sorption isotherms are fitted from (a) the BET model, (b) the GAB model, (c) the D&W model and (d) the GDW model. Experimental data points expressed as means +/- SD. Mann-Whitney tests performed between control and plasma data at each relative humidity (\*: p-value < 0.05; \*\*: p-value < 0.01). (e) Difference of water content between plasma and control conditions, considering the 4 aforementioned models.*

Before analyzing these models, it should be remembered that from a purely thermodynamic point of view, two main types of sorption processes can be distinguished: adsorption (seeds absorb moisture from the surrounding environment) and desorption (seeds lose moisture to the surrounding environment). In parallel, three distinct regions can be distinguished in sorption isotherms, based on the binding strength of water to the seed tissues [55], [56]. These regions can be distinguished considering several parameters: water content, type of sorption sites and nature of chemical bonds, as reported in **Table 1** and illustrated in **Figure 11**.

*Table 1. Comparative analysis of hydration dynamics in seed tissues by region-based water content levels.*

|  | Region I | Region II | Region III |
|---|---|---|---|
| **WC** | < 30 %$_{DW}$ | 30-80 %$_{DW}$ | > 80 %$_{DW}$ |
| **Seed sites** | Strong binding sites | Weak binding sites | Multimolecular sorption sites |
| **States** | Monolayer adsorption | Monolayer adsorption + Multilayer adsorption | Monolayer adsorption + Multilayer adsorption + Free water |
| **Chemical bonds** | Hydrogen bonds | Van der Waals bonds and additional hydrogen bonds | Capillary condensation and minimal van der Waals forces. |
| **Details** | Water molecules interact directly with the polar functional groups present in seed tissues, such as hydroxyl (-OH) groups in cellulose, or amine (-NH$_2$) groups and carbonyl (C=O) groups in proteins. These strong interactions result in a single layer (monolayer) of tightly bound water molecules on the surface of the seed tissue (**Figure 11**). | After formation of the monolayer at the strong binding sites, additional layers of water molecules can form on top, as shown in **Figure 11**. H$_2$O molecules are held in place not only by interacting with the seed tissue but also by hydrogen bonding to the already adsorbed water molecules from the monolayer. The forces holding these subsequent layers are weaker than the primary monolayer, so this water is more easily removed than that at the strong binding sites | Beyond the multilayer adsorption, as more water is adsorbed or as the tissue becomes more saturated, water starts to accumulate in larger quantities in the pores and cavities of the seed matrix. At high moisture content values, this water can lead to conditions favorable for microbial growth and other undesirable changes in the seed (**Figure 11**). |

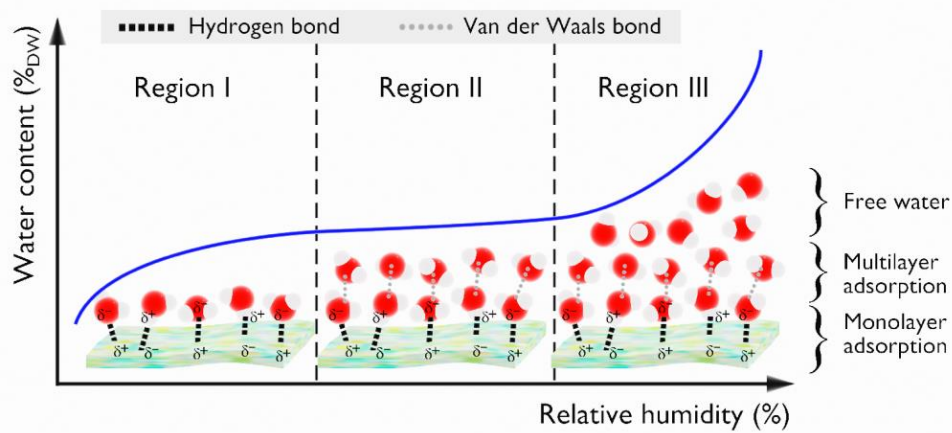

*Figure 11. Adsorption isotherm illustrating the equilibrium water content of Arabidopsis seeds as a function of different relative humidity levels.*

The biophysical parameters of the four models in **Figure 10** and the statistical coefficients assessing their fitting to the experimental data points ($R^2$) are reported in **Table 2**, considering control and plasma groups. The BET model, while fundamental to seed adsorption studies, only considers water adsorption in a single layer, which makes it relevant for low values of water affinity [57]. In contrast, the other three models consider seeds as complex, multilayered structures, in particular the GDW model which stands out for its global approach, integrating Langmuir-type adsorption on both inner and outer seed surfaces. In the BET model, the $M_m$ parameter is almost the same for the control and plasma groups. $M_m$ is the monolayer moisture content, i.e. the moisture considering that only a single layer of water molecules is adsorbed onto the seed surface (dimensionless quantity, normalised to 1). A low value of $M_m$ indicates that only a small amount of water is needed to saturate the monolayer on the seed surface. This could mean a smaller surface area or a lower affinity for water in the initial adsorption stage. It may suggest fewer sites on the seed surface for water binding or a less polar nature of the seed's surface. Conversely, a high value of $M_m$ indicates that the seed surface can hold a substantial amount of water before additional layers start to form. This could mean a larger surface area or higher affinity for water in the initial adsorption stage. It may also suggest that the seed has many polar sites or specific structures that can bond with water molecules effectively. Since here the $M_m$ parameter is almost the same for the control and plasma groups, this suggests that the plasma treatment does not change the strength of the binding sites. However, the $C_{BET}$ parameter is higher for the plasma group (16.0) than for control group (6.9). $C_{BET}$ is a dimensionless constant which relates to the energy of adsorption, i.e. a measure of the energy of interaction between the adsorbate molecules (water) and the seed surface. When $C_{BET}$ = 1, it implies that the energy of interaction between the adsorbed layers is the same as the liquid state of the adsorbate. However, this is a very rare scenario. Typically, $C_{BET}$ is greater than 1, and a higher value of $C_{BET}$ indicates stronger interactions between the first layer of adsorbed water molecules and the seed surface, compared to subsequent layers. In simpler terms, it indicates how tightly the first layer of water molecules binds to the seed. Since here the $C_{BET}$ parameter is 2.3 times higher for the plasma group than for the control group, this means that the adsorption is stronger for the plasma-treated seeds at low $H_{rel}$.

In the GAB model, the same findings apply to the $M_m$ parameter: the plasma treatment did not significantly change its value: $4.6 \times 10^{-2}$ and $4.6 \times 10^{-2}$ for control and plasma groups. This means that the plasma did not modify the amount of water on seed surface before additional layers start to form. Therefore, the plasma treatment is not responsible for larger surface areas and/or higher affinity for water in the initial adsorption stage. The $C_{GAB}$ parameter is a dimensionless constant, related to the enthalpy of sorption for the molecules in the multilayer. Here, its value is significantly increased, from 11.5 (Control) to 40.7 (Plasma). Before plasma exposure, the low value of $C_{GAB}$ suggests that the difference in adsorption energy between the initial monolayer and the subsequent layers is smaller. The interactions for multilayer adsorption are more similar to the monolayer adsorption, meaning there is less distinction in energy between the two. However, the higher value of $C_{GAB}$ obtained after plasma exposure indicates a large difference between the adsorption energy of the first layer (monolayer) and that of the subsequent layers. This means that thanks to the plasma treatment, the initial monolayer has a strong interaction with the seed and that once the monolayer is formed, the energy required to adsorb additional molecules in the subsequent layers is significantly lower. The GAB model is also interesting because it includes a third parameter which is totally ignored in the BET model. This parameter is K, a dimensionless constant that represents the ratio of the sorption enthalpy of water in the multilayer to that in the monolayer. It is a measure of how the strength of these interactions changes from the first layer of adsorbed water (the monolayer) to subsequent layers (multilayers). A scenario with K>1 is unlikely, as it would suggest an unusual situation where water-water interactions in the multilayers surpass those between water and the seed substrate in the monolayer. The situation where K = 1 would signify that the energy required for water sorption in the monolayer is equivalent to that in the multilayers. In our dataset, K is lower than 1, with values as low as $8.6 \times 10^{-3}$ and $8.6 \times 10^{-3}$ for control and plasma groups respectively. This means that the interactions between water molecules in the multilayers are weaker than the water-substrate interactions in the monolayer. This is typically the more common scenario since water molecules in the primary layer are directly bound to the substrate, resulting in stronger interactions. In contrast, water in the multilayers interacts more with other water molecules, leading to weaker overall interactions.

Although empirical, the D'Arcy & Watt model (D&W) aligns well with experimental data when the C parameter is zero. This suggests either an absence of weak sorption sites in the seeds or, if present, a water affinity that is negligible, as already underlined by Hay et al. [58]. The D&W model provides additional information compared with the two previous models, particularly with regard to the water affinity of sorption sites. For seeds from the control group, it is clear that water affinity is 4 times greater for strong sorption sites (K) than for multimolecular sites (k). Surprisingly, water affinity is totally negligible for weak sorption sites (C), as already underlined by Hay et al [58]. After exposure to cold plasma, weak sorption sites retain an unchanged affinity for water, while that of strong sorption sites increases significantly (from $4.74 \times 10^{-2}$ to $12.18 \times 10^{-2}$). The parameters K', C and k' act as intermediates from which the number of sorption sites per gram can be determined, differentiating between strong, weak and multimolecular sorption. Unsurprisingly, untreated seeds have a much higher number of strong sorption sites ($N_{strong} = 26.0 \times 10^{20}$ $g_{DW}^{-1}$) than multimolecular sorption sites ($N_{multi} = 5.35 \times 10^{20}$ $g_{DW}^{-1}$). The latter therefore play a negligible role compared with the former. Remarkably, cold plasma treatment reduces $N_{strong}$ to a value of $20.6 \times 10^{20}$ $g_{DW}^{-1}$, while $N_{multi}$ remains stable ($5.2 \times 10^{20}$ $g_{DW}^{-1}$). Consequently, the G&W model confirms that although cold plasma tends to reduce the number of strong sorption sites by 20.7 %, the overall affinity of the active sites has increased by over 150 %.

Finally, the GDW model indicates that the surface concentration of the primary adsorption sites ($a_{mL}$) is decreased by 29.5% after plasma exposure, which is consistent with the Mm parameters from BET and GAB models as well as with K' and $N_{strong}$ parameters from D&W model. If the plasma treatment slightly reduces the number of sorption sites, the other sites, especially the strong sorption sites, present a stronger binding between water and seed tissue, as highlighted by the value of the Langmuir constant which is increased by more than 280%, from $1.74 \times 10^{-5}$ Pa$^{-1}$ (Control) to $5.24 \times 10^{-5}$ Pa$^{-1}$ (Plasma). The GDW model provides also information on the proportion of water molecules that are bound to primary sites and that can transform into secondary sites. This proportion ($\omega$) is increased by 22.4% after plasma exposure. A similar kinetic constant (C) related to the adsorption on secondary sites was fitted for the two groups: no discrepancy was observed in terms of time scale of the adsorption mechanisms.

Table 2. Biophysical parameters of BET, GAB, D&W and GDW models and their values fitted on experimental water sorption isotherms, considering two groups of Arabidopsis dormant seeds: seeds without treatment (CTRL) and seeds after plasma exposure (PLASMA). *: dimensionless quantity; #: normalized to 0-1 range

| Model | | Parameters | Units | Ctrl | Plasma | (Pl-Ctrl)/Ctrl |
|---|---|---|---|---|---|---|
| BET | $M_m$ | Monolayer moisture content parameter | *, # | $5.2 \times 10^{-2}$ | $4.8 \times 10^{-2}$ | − 7.7 % |
| | $C_{BET}$ | Energy of adsorption constant | * | 6.9 | 16.0 | +131.9 % |
| | $R^2$ | Coeff. of determination | *, # | 0.9850 | 0.9940 | - |
| GAB | $M_m$ | Monolayer moisture content parameter | *, # | $4.6 \times 10^{-2}$ | $4.2 \times 10^{-2}$ | − 8.7 % |
| | $C_{GAB}$ | Sorption enthalpy for the molecules in the multilayer | * | 11.5 | 30.7 | +167.0 % |
| | K | Ratio of the sorption enthalpy of water in the multilayer to that in the monolayer | *, # | $8.6 \times 10^{-3}$ | $8.3 \times 10^{-3}$ | − 3.5 % |
| | $R^2$ | Coeff. of determination | *, # | 0.9970 | 0.9850 | - |
| D&W | K | Hydric affinity of the strong binding sites | * | $4.74 \times 10^{-2}$ | $12.18 \times 10^{-2}$ | +157.0 % |
| | K' | Scale factor relative to the number of strong binding sites | *, # | $7.79 \times 10^{-2}$ | $6.16 \times 10^{-2}$ | −20.9 % |
| | C | Strength and scale factor relative to the number of weak binding sites | *, # | $1.0 \times 10^{-9}$ | $1.0 \times 10^{-9}$ | 0 |
| | k | Water activity of multimolecular water | * | $1.02 \times 10^{-2}$ | $1.01 \times 10^{-2}$ | −1.0 % |
| | k' | Scale factor relative to the number of multimolecular sorption sites | *, # | $1.60 \times 10^{-2}$ | $1.56 \times 10^{-2}$ | −2.5 % |
| | $N_{strong}$ | Number of strong sorption sites per gram | $\times 10^{20}\ g_{DW}^{-1}$ | 26.0 | 20.6 | −20.8 % |
| | $N_{weak}$ | Number of weak sorption sites per gram | $\times 10^{20}\ g_{DW}^{-1}$ | Negligible | Negligible | - |
| | $N_{multi}$ | Number of multimolecular sorption sites per gram | $\times 10^{20}\ g_{DW}^{-1}$ | 5.35 | 5.22 | −2.4 % |
| | $R^2$ | Coeff. of determination | *, # | 0.9991 | 0.9991 | - |
| GDW | $a_{mL}$ | Surface concentration of primary Langmuir-type adsorption sites | $mol.g_{DW}^{-1}$ | $8.8 \times 10^{-2}$ | $6.2 \times 10^{-2}$ | −29.5 % |
| | $K_L$ | Langmuir constant | $Pa^{-1}$ | $1.74 \times 10^{-5}$ | $5.24 \times 10^{-5}$ | +282.3 % |
| | ω | Proportion of water molecules bound to primary sites that subsequently transform into secondary sites | *, # | $2.2 \times 10^{-1}$ | $2.7 \times 10^{-1}$ | +22.7 % |
| | C | Kinetic constant related to the adsorption on the secondary sites | *, # | $1.0 \times 10^{-2}$ | $1.0 \times 10^{-2}$ | 0.0 % |
| | $a_{prim}$ | Primary site adsorption value | $mol.g_{DW}^{-1}$ | $3.3 \times 10^{-3}$ | $6.7 \times 10^{-3}$ | +103.0 % |
| | $R^2$ | | *, # | 0.9991 | 0.9995 | 0.0 % |

The four models provide a better understanding of the biophysical parameters of seed hydration, in particular how seeds interact with water at the molecular level and how this interaction can be modified by plasma treatment. The results show that plasma treatment does not significantly alter the initial monolayer of moisture on the seed surface, but does intensify the force of interaction between water and seed. In the D&W model, it is evident that plasma treatment increases the affinity of water for strong sorption sites, while weak sorption sites remain largely unchanged. In addition, the GDW model highlights a decrease in primary adsorption sites after plasma exposure, but emphasizes a stronger binding between water and seed tissue. In essence, while plasma treatment may slightly reduce the number of sorption sites, it significantly amplifies the overall affinity of water. This result is in line with the hypothesis that plasma-generated reactive oxygen species play a role in the incorporation of oxygen-containing groups into the internal structures of the seed. These oxygen-containing groups make the seed tissues more hydrophilic, thereby increasing their affinity for water, thus possibly promoting the disruption of seed dormancy.

## 4. Conclusion

In this article, we have explored the bidirectional interactions between plasma and seeds, in particular how the electrical properties of plasma are influenced by seed feedback. When dormant Arabidopsis seeds of varying water content (3 %$_{DW}$, 10 %$_{DW}$ and 30 %$_{DW}$) are subjected to C2AP, we observe that an increase in water content increases the capacitive current of the DBD. This increase also results in a greater number of low-energy streamers, which preferentially interact with the seeds rather than the dielectric barrier or counter-electrode. Such interactions are thought to be more effective in enhancing seed germination, including the release of dormancy.

To elucidate the implications of these interactions on the physical properties of the seeds, LCR measurements have been carried out before and after plasma exposure. These measurements reveal that WC notably affects dielectric losses at frequencies below 1 kHz, which is attributed to relaxation of the polarization of water molecules. Despite this, C2AP does not significantly alter the dielectric parameters of seeds, preserving their starch, protein and lipid content. However, it may alter some overall seed properties, probably due to the porosity of the seeds, which allows plasma-generated reactive oxygen species to penetrate and potentially graft themselves as oxygen groups inside the seeds.

Finally, we have investigated the affinity between water and seed and its modification by plasma treatment. Water sorption isotherms fitted to thermodynamic models such as the generalized model of D'Arcy and Watt demonstrate that C2AP mainly strengthens the link between water molecules and the seed by modifying molecular interactions rather than the existing moisture layer of the seed. This suggests an increase in overall seed hydrophilicity after plasma treatment, although there may be a reduction in the number of water adsorption sites. This increased hydrophilicity would result from the incorporation of oxygen-containing groups that plasma delivers into the internal seed structures. This plasma-induced surface modification may play a role in releasing Arabidopsis seed dormancy.

# Acknowledgements

This work was supported by a PhD grant from Sorbonne Université and received financial state aid as part of the PF2ABIOMEDE platform co-funded by « Région Ile-de-France » (Sesame, Ref. 16016309) and Sorbonne Université (technological platforms funding).